\documentclass[conference,a4paper,onecolumn]{IEEEtran}

\usepackage[utf8]{inputenc} 
\usepackage[T1]{fontenc}
\usepackage{url}              % provides \url{...}
\usepackage{cite}             % improves presentation of citations
\usepackage[cmex10]{amsmath}  % Use the [cmex10] option to ensure complicance
\usepackage{mleftright}       % fix to wrong spacing of \left-,
\mleftright                   % \middle- \right-commands 
\usepackage{graphicx}         % provides \includegraphics{...} to
\usepackage{booktabs}         % fixes poor spacing in tables and

\usepackage[bookmarks]{hyperref}
\usepackage{amssymb}
\usepackage{amsmath}
\usepackage{amsthm}
\usepackage[abs]{overpic}
\usepackage{tcolorbox}
\usepackage{color,colordvi}
\usepackage{bbm}
\usepackage{cleveref}
\usepackage{stmaryrd}
\usepackage{dsfont}
\usepackage{mathtools}

\usepackage{centernot}
\usepackage{tikz}
\usepackage{pgfplots}

\def\openone{\leavevmode\hbox{\small1\kern-3.8pt\normalsize1}}

\def\11{\mathbb{I}}

\def\Ppg{P_{\operatorname{guess}}^{\operatorname{pg}}}

\newtheorem{definition}{Definition}[section]
\newtheorem{proposition}[definition]{Proposition}
\newtheorem{lemma}[definition]{Lemma}

\newtheorem{theorem}[definition]{Theorem}

\newtheorem{conjecture}[definition]{Conjecture}

\newcommand{\tr}{\mathop{\rm Tr}\nolimits}

\usepackage{pst-node}
\usepackage{tikz-cd}

\newcommand{\etal}{\textit{et al. }}

\newcommand{\Renyi}{R{\'e}nyi~}

\usepackage{graphicx}
\usepackage{setspace}
\usepackage{verbatim}
\usepackage{subfig}

\DeclareFontFamily{U}{mathx}{\hyphenchar\font45}
\DeclareFontShape{U}{mathx}{m}{n}{<-> mathx10}{}
\DeclareSymbolFont{mathx}{U}{mathx}{m}{n}
\DeclareMathAccent{\widebar}{0}{mathx}{"73}

\numberwithin{equation}{section}

\DeclareRobustCommand\openone{\leavevmode\hbox{\small1\normalsize\kern-.33em1}}

\newcommand{\be}{\begin{equation}}
	\newcommand{\ee}{\end{equation}}
\newcommand{\bea}{\begin{eqnarray}}
	\newcommand{\eea}{\end{eqnarray}}
\newcommand{\beas}{\begin{eqnarray*}}
	\newcommand{\eeas}{\end{eqnarray*}}

\begin{document}

\title{Chain Rules for \Renyi Information Combining}
\author{
\IEEEauthorblockN{Christoph Hirche\IEEEauthorrefmark{1}\IEEEauthorrefmark{2}, Xinyue Guan\IEEEauthorrefmark{3}, Marco Tomamichel\IEEEauthorrefmark{2}\IEEEauthorrefmark{4}}
\IEEEauthorblockA{\IEEEauthorrefmark{1}Zentrum Mathematik, Technical University of Munich, 85748 Garching, Germany}
\IEEEauthorblockA{\IEEEauthorrefmark{2}Centre for Quantum Technologies, National University of Singapore, Singapore}
\IEEEauthorblockA{\IEEEauthorrefmark{3}River Valley High School, Singapore}
\IEEEauthorblockA{\IEEEauthorrefmark{4}Department of Electrical and Computer Engineering, National University of Singapore, Singapore}}
%\author{%
 % \IEEEauthorblockN{Anonymous Authors}
%  \IEEEauthorblockA{%
 %   Please do NOT provide authors' names and affiliations\\
  %  in the paper submitted for review, but keep this placeholder.\\
 %   ISIT23 follows a \textbf{double-blind reviewing policy}.}
%}

\maketitle

\begin{abstract}
Bounds on information combining are a fundamental tool in coding theory, in particular when analyzing polar codes and belief propagation. They usually bound the evolution of random variables with respect to their Shannon entropy. In recent work this approach was generalized to \Renyi $\alpha$-entropies. However, due to the lack of a traditional chain rule for \Renyi entropies the picture remained incomplete. In this work we establish the missing link by providing \Renyi chain rules connecting different definitions of \Renyi entropies by Hayashi and Arimoto. This allows us to provide new information combining bounds for the Arimoto \Renyi entropy. 
In the second part, we generalize the chain rule to the quantum setting and show how they allow us to generalize results and conjectures previously only given for the von Neumann entropy. In the special case of $\alpha=2$ we give the first optimal information combining bounds with quantum side information. 
\end{abstract}

\section{Introduction} 

Many tasks in information theory are concerned with the evolution of random variables and their corresponding entropies under certain ``combining operations''. 
A particularly relevant example of such an operation is the addition of two independent random variables (with values in some group). In this case, the entropy can be easily computed since we know that the addition of two random variables has a probability distribution which corresponds to the convolution of the probability distributions of the individual random variables. 
The picture changes when we have random variables with \emph{side information}. Now we are interested in the entropy of the sum conditioned on all the available side information. Evaluating this is substantially more difficult. The area of \textit{bounds on information combining} is concerned with finding optimal entropic bounds on the resulting conditional entropy. 

A particularly important case is that of two binary random variables with side information of arbitrary finite dimension. This is the setting we will consider in this work as it has many applications e.g. in coding theory for polar codes~\cite{AT14} and Shannon theory~\cite{GKBook, RU08}. 
An optimal lower bound for the resulting Shannon entropy was given by Wyner and Ziv in \cite{WZ73}, the well known \textit{Mrs.~Gerber's Lemma}. 
Following this result, additional approaches to the problem have been found which also led to an \emph{upper} bound on the conditional entropy of the combined random variables. One proof method and several additional applications can be found e.g.~in \cite{RU08} along with the optimal upper bound.

In this work, we discuss the extension of these information combining bounds to R\'enyi entropies. While the unconditional R\'enyi entropy is clearly defined, one can find many different definitions of the conditional R\'enyi entropy in the literature. We focus on two, $H_\alpha^H$ and $H_\alpha^A$, which show particularly favorable properties. Recently, optimal bounds on
\begin{align}
H_\alpha^*(X_1+X_2|Y_1Y_2)
\end{align}
were shown in~\cite{hirche2020renyi}. This corresponds to the entropy of the check node in belief propagation or the \textit{worse} channel in polar coding. For many applications, e.g.~in coding theory, one however also needs bounds on the complementary entropy
\begin{align}
H_\alpha^*(X_2|X_1+X_2,Y_1Y_2), 
\end{align}
which corresponds to the variable node or \textit{better} channel. 
In the Shannon entropy case, i.e. $\alpha=1$, one can use the well known chain rule for conditional entropy to translate bounds on the first entropy into bounds on the second. In general however, no such nice chain rules hold for the considered \Renyi entropies. The central subject of this work are equalities that can take the role of a chain rule for information combining. 
As we will see, these \Renyi entropy chain rules directly connect $H_\alpha^H$ and $H_\alpha^A$ in the sense that one can directly translate bounds between
\begin{align}
H_\alpha^H(X_1+X_2|Y_1Y_2) \quad \textnormal{and} \quad H_\alpha^A(X_2|X_1+X_2,Y_1Y_2).
\end{align}
This might seem like a surprising connection at first, however we will afterwards see that it appears more natural in the framework of quantum information theory. 

The theme of bounds on information combining was generalized to include quantum side information in~\cite{hirche2018bounds}. In~\cite{hirche2018bounds} non-optimal lower bounds on $H(X_1+X_2|Y_1Y_2)$ are given and an application to quantum polar codes~\cite{wilde2012polar, renes2014polar, hirche2015polar} is discussed. Additionally, a conjecture was given that the optimal lower and upper bounds are a symmetrized version of the classical bounds. Combining the chain rule for the von Neumann entropy with the concept of dual channels~\cite{renes2014polar,renes2015alignment,renes2017duality} it turns out that this symmetry is actually an essential feature of any optimal von Neumann bound in the quantum setting. 
The quantum setting is intrinsically more difficult as we do not have a good general understanding of how to handle conditioning on quantum side information. 

An added difficulty in generalizing the previous results to \Renyi entropies is that while we already know of several classical versions of conditional \Renyi entropies there are vastly more options in the quantum setting. 
In this work we will mainly consider the extensions $\widetilde H_\alpha^\downarrow$ and $\widebar H_\alpha^\uparrow$ and will see that they obey a number of desirable properties including a quantum generalization of our \Renyi entropy chain rule. Subsequently, several properties of the von Neumann case carry over to the \Renyi setting, including symmetry of the optimal bounds and the connection
\begin{align}
\widetilde H_\alpha^\downarrow(X_1+X_2|Y_1Y_2) \,\longleftrightarrow\, \widebar H_\alpha^\uparrow(X_2|X_1+X_2,Y_1Y_2).
\end{align}
Finally, we provide optimal bounds for some special cases and give a conjecture for optimal bounds for general $\alpha$. In the appendix we discuss numerical evidence, also concerning other \Renyi conditional entropies. Also all technical proofs are given in the appendix. 

%In all cases, we find the optimal lower and upper bounds given a certain convexity or concavity property of an associated function. In many cases we will also give the ranges of $\alpha$ for which the desired property holds. Furthermore, we identify some remarkable cases where the bounds hold with equality. These results are briefly summarized in Figure~\ref{Fig:Results}. 
%As in the Shannon entropy case, our bounds are optimal in the sense that they are attained by the binary symmetric channel and the binary erasure channel with equality. 

%Finally, we briefly discuss the application of our results to channel polarization. We show that our results can easily be combined with a simplified proof strategy for polarization from~\cite{AT14} to show polarization of $H_\alpha^J$ under the polar coding transformation, therefore reproducing a result previously found in~\cite{zheng2019polarization}.

\bigskip
\noindent\textbf{Previous results:} Before starting we will briefly recall some central results from~\cite{hirche2020renyi}. These will serve as a point of comparison but also be needed as tools later on. For simplicity we state here a combination of~\cite[Theorem IV.8 \& Lemma IV.10]{hirche2020renyi} for $H_\alpha^H$ which will later be defined precisely in~\eqref{Eq:HHa-def}. In the following, we denote the binary \Renyi entropy as $h_\alpha$.

 \begin{theorem}[$H_{\alpha}^{H}$ BSC-bound~\cite{hirche2020renyi}]\label{thm:old-H-BSC}
For $\alpha\in(0,2]\cup[3,\infty)$, 
 \begin{align}
 &H_\alpha^H(X_1+X_2|Y_1Y_2) \geq h_\alpha(h_\alpha^{-1}( H_\alpha^H(X_1|Y_1))\ast h_\alpha^{-1}( H_\alpha^H(X_2|Y_2))). \label{BSCdownH}
 \end{align}
For $\alpha\in[2,3]$ the same holds with $\geq$ exchanged by $\leq$. Note that for $\alpha \in \{2,3\}$ the above holds with equality. These bounds are tight in the sense that equality is achieved by binary symmetric channels. 
 \end{theorem}
We define $\delta_\alpha :=2^{1-\alpha}$ and $K_\alpha^*(X|Y) :=e^{(1-\alpha)H_\alpha^*(X|Y)}$. By combining~\cite[Theorem IV.9 \& Lemma IV.10]{hirche2020renyi} we get the following. 
\begin{theorem}[$H_{\alpha}^{H}$ BEC-bound~\cite{hirche2020renyi}]\label{thm:old-H-BEC}
For $\alpha\in(0,2]\cup[3,\infty)$,  
\begin{align}
 &H_\alpha^H(X_1+X_2|Y_1Y_2) \leq \frac{1}{1-\alpha}\log\frac{(\delta_\alpha - K_\alpha^H(X_1|Y_1))(\delta_\alpha - K_\alpha^H(X_2|Y_2))}{1-\delta_\alpha} + \delta_\alpha .
 \end{align}
For $\alpha\in[2,3]$ the same holds with $\geq$ exchanged by $\leq$. Note that for $\alpha \in \{2,3\}$ the above holds with equality. These bounds are tight in the sense that equality is achieved by binary erasure channels. 
\end{theorem}
Similar results hold for the Arimoto conditional entropy and can also be found in~\cite{hirche2020renyi}.

\section{Classical \Renyi entropies}

In this section we will formally define the considered classical \Renyi entropies and give some important properties. The main definitions are, 
\begin{align}
H^H_\alpha(X|Y)_p &:= \frac{1}{1-\alpha}\log\sum_y\sum_x p(y) p(x|y)^\alpha \label{Eq:HHa-def}\\
H^A_\alpha(X|Y)_p &:= \frac{\alpha}{1-\alpha}\log\sum_y p(y) \left(\sum_x p(x|y)^\alpha\right)^\frac{1}{\alpha}, 
\end{align}
where the first was first discussed by Hayashi and Skoric \etal~\cite{Hayashi11, Skoric11} and the second by Arimoto~\cite{Arimoto77}. 
For the latter we often invert $\alpha$,
\begin{align}
H^A_{\frac{1}{\alpha}}(X|Y)_p = \frac{1}{\alpha-1}\log\sum_y p(y) \left(\sum_x p(x|y)^\frac{1}{\alpha}\right)^{\alpha}. 
\end{align}
It holds that, 
\begin{align}
\lim_{\alpha\rightarrow 1} H^*_\alpha(X|Y) = H(X|Y). 
\end{align}
There are several more definitions of conditional \Renyi entropy in the literature, for some of which information combining bounds have been considered~\cite{hirche2020renyi}. However, the ones we choose here are the only of those that obey monotonicity, 
\begin{align}
     H^*_\alpha(X|YZ) \leq H^*_\alpha(X|Z).
\end{align}
Unfortunately, neither of these definitions obeys a traditional chain rule, meaning in general
 \begin{align}
 H^*_\alpha(X|YZ) \neq H^*_\alpha(XY|Z) - H^*_\alpha(Y|Z).
 \end{align}
However, when we consider information combining the following observation can be shown to take the place of a traditional chain rule. 
\begin{lemma}[Classical \Renyi chain rule]\label{lem:cl-CR}
Let $p(x,y)$ and $\bar p(x,y)$ be joint probability distributions related by, 
\begin{align}
\bar p(x,y) = \frac{p(x|y)^\alpha p(y) }{\sum_{x,y} p(x|y)^\alpha p(y)}, \label{Eq:Pxy-relation}
\end{align}
then 
\begin{align}
H^H_\alpha(X|Y)_p = H^A_{\frac{1}{\alpha}}(X|Y)_{\bar p}. 
\end{align}
Similarly, let 
\begin{align}
\hat p(x,y,z) = \frac{p(z) p(x,y|z)^\alpha}{\sum_{x,y,z}p(z) p(x,y|z)^\alpha }, 
\end{align}
then
\begin{align}
H^H_\alpha(XY|Z)_p &= H^A_{\frac{1}{\alpha}}(X|YZ)_{\hat p} + H^H_\alpha(Y|Z)_p \\
&= H^A_{\frac{1}{\alpha}}(XY|Z)_{\hat p}.
\end{align}
\end{lemma}
Note that the relations in the previous lemma are bijections. For example the inverse of Equation~\eqref{Eq:Pxy-relation} is given by
\begin{align}
p(x,y) = \bar p(x|y)^{\frac{1}{\alpha}} \bar p(y) \frac{\left( \sum_x \bar p(x|y)^{\frac1{\alpha}}\right)^{\alpha-1}}{\sum_y \bar p(y) \left( \sum_x \bar p(x|y)^{\frac1{\alpha}}\right)^{\alpha} }. 
\end{align}
%This allows us to use the lemma starting either from $p$ or $\bar p$. 
%\MT{This paragraph is a bit pedestrian. I think you just want to say that (2.7) is a bijection and the reverse map is given by this?}

Maybe surprisingly these new rules combine both types of considered conditional entropy. In the next section we will discuss their application to information combining. Afterwards, when considering quantum side information, we will see that these two entropies are in fact dual to each other in a certain sense, giving a natural connection in that more general setting. 

\section{\Renyi information combining bounds}

We start by making the considered setting precise. We have two joint probability distributions $p_1(x_1,y_1)$ and $p_2(x_2,y_2)$, which we associate with channels 
\begin{align}
W_1 &\sim p_1(x_1,y_1), \\
W_2 &\sim p_2(x_2,y_2), 
\end{align}
leaving the input implicit, and we often use the notation
\begin{align}
H^*_\alpha(W_i) = H^*_\alpha(X_i|Y_i)_{p_i}.
\end{align}
Here, we consider information combining via XOR operation while keeping a copy of $X_2$. Denote $Z = X_1+X_2$, then the probability after combining is
\begin{align}
p(z,x_2,y_1,y_2) = p_1(z\oplus x_2,y_1) p_2(x_2,y_2). 
\end{align}
From this distribution we can synthesize two different combined channels with associated entropies
\begin{align}
H^*_\alpha(W_1 \boxast W_2) &= H^*_\alpha(X_1+X_2|Y_1Y_2)_p \\
H^*_\alpha(W_1 \varoast W_2) &= H^*_\alpha(X_2|X_1+X_2,Y_1Y_2)_p.
\end{align}
These synthesized channels have a central role e.g. for the well known polar codes~\cite{arikan2009channel}. 
To asses the change in entropy resulting from these channel combinations we are interested in bounds on the above quantities in terms of the original $H^*_\alpha(W_i)$. Such bounds are well known in the case $\alpha=1$. Additionally, bounds on $H^*_\alpha(W_1 \boxast W_2)$ were recently proven in~\cite{hirche2020renyi} for several different choices of $H^*_\alpha$. 

For $\alpha=1$ bounds on $H(W_1 \varoast W_2)$ follow easily from those on $H(W_1 \boxast W_2)$ due to the well known chain rule. Namely, we have
\begin{align}
&H(X_2|X_1+X_2,Y_1Y_2) + H(X_1+X_2|Y_1Y_2) \\
&= H(X_1+X_2,X_2|Y_1Y_2) \\
&= H(X_1,X_2|Y_1Y_2) = H(X_1|Y_1) + H(X_2|Y_2), 
\end{align}
which can be equivalently written as 
\begin{align}
H(W_1 \varoast W_2) + H(W_1 \boxast W_2) = H(W_1) + H(W_2).
\end{align}
The chain rule also implies that the overall entropy is preserved under combining operations such as polar codes which is a crucial property to show that polar codes are capacity achieving~\cite{arikan2009channel}. 
Such clean chain rules however generally do not hold when replacing the Shannon entropy with \Renyi entropies. We can however use the new chain rules from the previous section instead. 

\begin{lemma}[Information combining chain rule]\label{Lem:CICCR}
Given channels $W_1$ and $W_2$, there exist channels $W'_1$ and $W'_2$ with $H^H_\alpha(W_i) =  H^A_{\frac{1}{\alpha}}(W'_i)$ such that 
\begin{align}
H^H_\alpha(W_1) + H^H_\alpha(W_2) = H^A_{\frac{1}{\alpha}}(W'_1\varoast W'_2) + H^H_\alpha(W_1\boxast W_2).
\end{align}
More precisely, these channels are given by
\begin{align}
W'_i\sim \bar p_i(x_i,y_i) = \frac{p_i(x_i|y_i)^\alpha p_i(y_i) }{\sum_{x,y} p_i(x_i|y_i)^\alpha p_i(y_i)}.
\end{align}
\end{lemma}
With this new chain rule we can prove new information combining bounds for the Arimoto \Renyi entropy. Remarkably, the proof uses the Hayashi \Renyi entropy, but the final result is expressed only in terms of Arimotos entropy. 
\begin{theorem}\label{Thm:AclUpper}
    For $\alpha\in(0,\frac13]\cup[\frac12,\infty)$, it holds that
    \begin{align}
    &H^A_{\alpha}(X_2|X_1+X_2,Y_1Y_2) \leq H^A_{\alpha}(X_1|Y_1) + H^A_{\alpha}(X_2|Y_2) - h_{\frac1{\alpha}}(h_{\frac1{\alpha}}^{-1}( H_{\alpha}^A(X_1|Y_1))\ast h_{\frac1{\alpha}}^{-1}( H_{\alpha}^A(X_2|Y_2))).
    \end{align}
    For $\alpha\in[\frac13,\frac12]$ the same holds with $\leq$ exchanged by $\geq$. Note that for $\alpha \in \{\frac13,\frac12 \}$ the above holds with equality. These bounds are tight in the sense that equality is achieved by binary symmetric channels. 
\end{theorem}

For the next result recall $\delta_{\frac1{\alpha}}=2^{\frac{\alpha-1}{\alpha}}$ and $K_{\frac1{\alpha}}^*(X|Y) :=e^{\frac{\alpha-1}{\alpha}H_\alpha^*(X|Y)}$. 

\begin{theorem}\label{Thm:AclLower}
   For $\alpha\in(0,\frac13]\cup[\frac12,\infty)$, it holds that
    \begin{align}
        &H^A_{\alpha}(X_2|X_1+X_2,Y_1Y_2) \geq H^A_{\alpha}(X_1|Y_1) + H^A_{\alpha}(X_2|Y_2) - \frac{\alpha}{\alpha-1}\log\frac{(\delta_{\frac1{\alpha}} - K_{\frac1{\alpha}}^A(X_1|Y_1))(\delta_{\frac1{\alpha}} - K_{\frac1{\alpha}}^A(X_2|Y_2))}{1-\delta_{\frac1{\alpha}}} + \delta_{\frac1{\alpha}}.
    \end{align}
    For $\alpha\in[\frac13,\frac12]$ the same holds with $\geq$ exchanged by $\leq$. Note that for $\alpha \in \{\frac13,\frac12 \}$ the above holds with equality. These bounds are tight in the sense that equality is achieved by binary erasure channels. 
\end{theorem}

\section{Quantum \Renyi entropies}

We now discuss the extension of the above results to the quantum setting. Essentially, quantum states $\rho_A$, i.e. normalized positive semidefinite matrices, take the role of probabilities. Now, consider the following entropies, see e.g.~\cite{tomamichel2015quantum},
\begin{align}
\widetilde H^\downarrow_\alpha(A|B)_\rho &= \frac{1}{1-\alpha}\log\tr\left(\rho_B^{\frac{1-\alpha}{2\alpha}}\rho_{AB}\rho_B^{\frac{1-\alpha}{2\alpha}}\right)^\alpha \\
\widebar H^\uparrow_\alpha(A|B)_\rho &= \frac{\alpha}{1-\alpha}\log\tr\left( \left(\tr_A\rho_{AB}^\alpha\right)^{\frac{1}{\alpha}}\right).  
\end{align}
The two entropies are connected by a duality relation~\cite{tomamichel2014relating}, stating that for pure states $\Psi_{ABC}$, 
\begin{align}
\widetilde H^\downarrow_\alpha(A|B)_\Psi = - \widebar H^\uparrow_{1/\alpha}(A|C)_\Psi. \label{Eq:H-duality}
\end{align}
Furthermore, for classical states $\rho_{XY}$ these entropies reduce to those discussed in the previous sections, 
\begin{align}
\widetilde H^\downarrow_\alpha(X|Y)_\rho &=  H^H_\alpha(X|Y)_p,   \\
\widebar H^\uparrow_\alpha(X|Y)_\rho &= H^A_\alpha(X|Y)_p. 
\end{align}
Similar to the classical case, no traditional chain rule holds. However, we can show the following when both entropies are allowed to appear in the chain rule. 
\begin{lemma}[Quantum \Renyi chain rule]\label{Lem:QCR}
Let $\rho_{AB}$ and $\bar\rho_{AB}$ be quantum states related by
\begin{align}
\bar\rho_{AB} = \frac{\left(\rho_B^{\frac{1-\alpha}{2\alpha}}\rho_{AB}\rho_B^{\frac{1-\alpha}{2\alpha}}\right)^\alpha}{\tr\left(\rho_B^{\frac{1-\alpha}{2\alpha}}\rho_{AB}\rho_B^{\frac{1-\alpha}{2\alpha}}\right)^\alpha}, 
\end{align}
then
\begin{align}
\widetilde H^\downarrow_\alpha(A|B)_\rho = \widebar H^\uparrow_{\frac{1}{\alpha}}(A|B)_{\bar\rho}.
\end{align}
Furthermore, if 
\begin{align}
\widetilde\rho_{ABC} = \frac{\left(\rho_C^{\frac{1-\alpha}{2\alpha}}\rho_{ABC}\rho_C^{\frac{1-\alpha}{2\alpha}}\right)^\alpha}{\tr\left(\rho_C^{\frac{1-\alpha}{2\alpha}}\rho_{ABC}\rho_C^{\frac{1-\alpha}{2\alpha}}\right)^\alpha}, 
\end{align}
then
\begin{align}
\widetilde H^\downarrow_\alpha(AB|C)_\rho = \widebar H^\uparrow_{\frac{1}{\alpha}}(A|BC)_{\widetilde\rho} + \widetilde H^\downarrow_\alpha(B|C)_\rho = \widebar H^\uparrow_{\frac{1}{\alpha}}(AB|C)_{\widetilde\rho}.
\end{align}
\end{lemma}
%\begin{proof}
%Similarly to the classical case, this can be shown by substitution. For details, see Appendix~\ref{Appendix:Proofs}. 
%\end{proof}
Another \Renyi generalization of the chain rule was recently proven in~\cite[Theorem 3.2]{dupuis2020entropy} where the authors show
\begin{align}
\widetilde H^\downarrow_\alpha(AB|C)_\rho = \widetilde H^\downarrow_\alpha(A|BC)_{\widehat\rho} + \widetilde H^\downarrow_\alpha(B|C)_\rho,
\end{align} 
with $\widehat\rho$ defined in~\cite[Equation (13)]{dupuis2020entropy}. However, in this formulation it is not known if $\widehat\rho$ represents a meaningful information combining operation as it does in our chain rule. 

\section{Information combining with quantum side information}

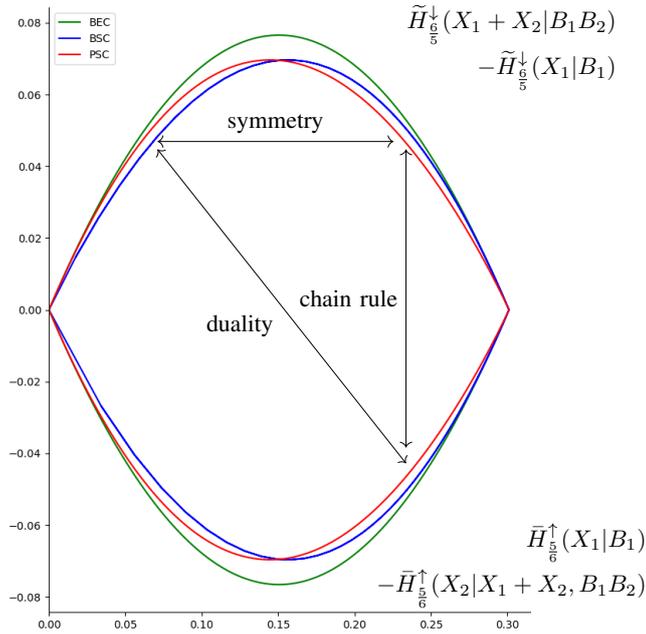
\begin{figure}[t]
\centering
\scalebox{0.9}{\begin{tikzpicture}
    \node[anchor=south west,inner sep=0] (image) at (0,0) {\includegraphics[width=0.5\textwidth]{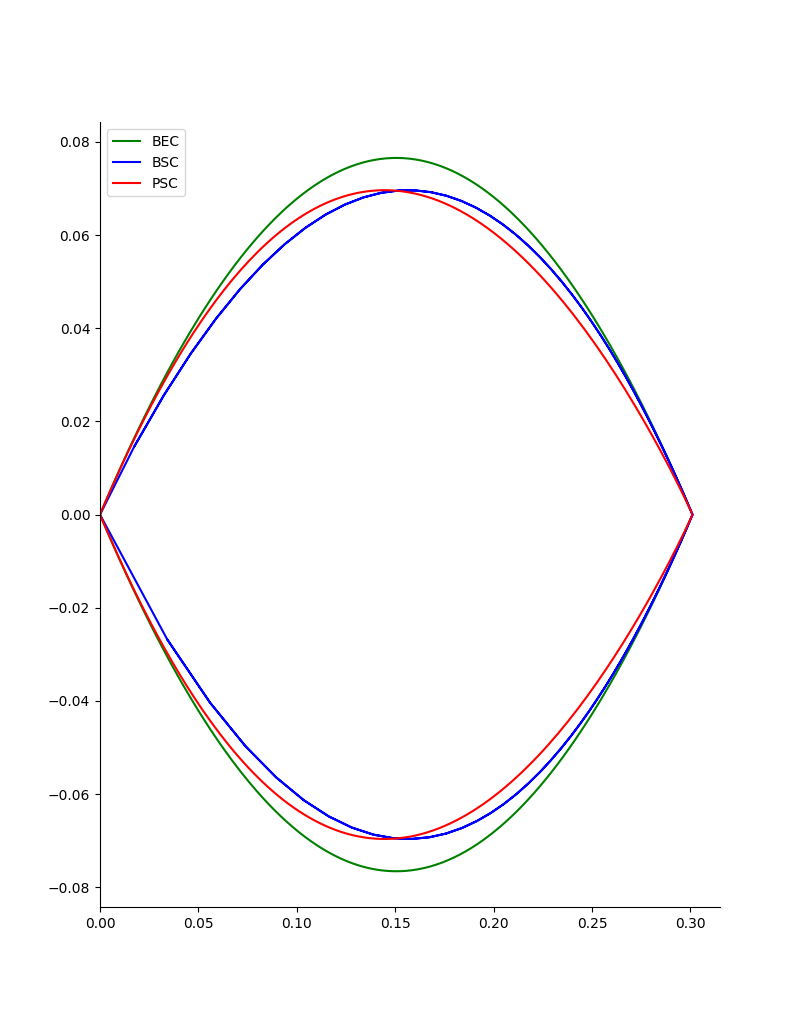}};
    \begin{scope}[x={(image.south east)},y={(image.north west)}]
        \draw[<->] (0.3,0.71) to ["symmetry"] (0.68,0.71) ;
        \draw[<->] (0.7,0.32) to ["chain rule"] (0.7,0.7) ;
        \draw[<->] (0.7,0.3) to ["duality"] (0.3,0.7) ;
        \node[] at (0.87,0.83){$\begin{aligned}\widetilde H_{\frac65}^\downarrow(X_1+X_2|B_1B_2) \\
        -\widetilde H_{\frac65}^\downarrow(X_1|B_1)\end{aligned}$};
        \node[] at (0.87,0.17){$\begin{aligned}\widebar H_{\frac56}^\uparrow(X_1|B_1) \\ - \widebar H_{\frac56}^\uparrow(X_2|X_1+X_2,B_1B_2)\end{aligned}$};
    \end{scope}
\end{tikzpicture}}
\caption{\label{Fig:relations} Setting with quantum side information. This figure shows the three rules that translate between points for which probability distributions exist: Duality, symmetry and the chain rule. Any 2 can be used to prove the third. The plotted graphs are the binary erasure channel (BEC), binary symmetric channel (BSC) and pure state channel (PSC) bounds for two different information combining entropies that are dual to each other. We consider $W_1=W_2$ and on the x-axes we plot over the individual entropy before combining.}
\end{figure}

We stick with the general setting of~\cite{hirche2018bounds}. In short, we associate classical-quantum channels with classical-quantum states
\begin{align}
W_1 &\sim \rho_1^{X_1B_1} \\
W_2 &\sim \rho_2^{X_2B_2}, 
\end{align}
where 
\begin{align}
\rho^{X_iB_i}_i = \frac12 |0\rangle\langle 0| \otimes \sigma_0^{B_i} + \frac12 |1\rangle\langle 1| \otimes \sigma_1^{B_i}
\end{align}
and we often use the notation
\begin{align}
H_\alpha(W_i) = H_\alpha(X_i|B_i)_{\rho_i}.
\end{align}
After applying a CNOT gate to the classical systems we have the joint state
\begin{align}
&\tau^{(X_1+X_2)X_2B_1B_2} \\
&= \sum_{z,x_2} \frac14 |z\oplus x_2\rangle\langle z\oplus x_2| \otimes |x_2\rangle\langle x_2| \otimes \sigma^{B_1}_z \otimes \sigma^{B_2}_{x_2}. 
\end{align}

We start by showing a chain rule for information combining. 
\begin{lemma}[Information combining chain rule]\label{Lem:QICCR}
Given channels $W_1$ and $W_2$, there exist channels $W'_1$ and $W'_2$ with $\widetilde H^\downarrow_\alpha(W_i) = \widebar H^\uparrow_{\frac{1}{\alpha}}(W'_i)$ such that 
\begin{align}
\widetilde H^\downarrow_\alpha(W_1) + \widetilde H^\downarrow_\alpha(W_2) = \widebar H^\uparrow_{\frac{1}{\alpha}}(W'_1\varoast W'_2) + \widetilde H^\downarrow_\alpha(W_1\boxast W_2).
\end{align}
More precisely, these channels are given by
\begin{align}
W'_i\sim \bar\rho_{X_iB_i} = \frac{\left(\rho_{B_i}^{\frac{1-\alpha}{2\alpha}}\rho_{X_iB_i}\rho_{B_i}^{\frac{1-\alpha}{2\alpha}}\right)^\alpha}{\tr\left(\rho_{B_i}^{\frac{1-\alpha}{2\alpha}}\rho_{X_iB_i}\rho_{B_i}^{\frac{1-\alpha}{2\alpha}}\right)^\alpha}.
\end{align}
\end{lemma}

We will also need the concept of duality discussed in~\cite{renes2014polar,renes2015alignment,renes2017duality}. Basically, for every $W$ there is a $W^\bot$ with $H(W) = \log2 - H^\bot(W^\bot)$, where $H^\bot$ is the dual entropy to $H$ in the sense of Equation~\eqref{Eq:H-duality}. In particular, one then also has $H(W_1\boxast W_2) = \log2 - H^\bot(W_1^\bot\varoast W_2^\bot)$. For a summary of duality properties in the context of information combining, see also~\cite{hirche2018bounds}. 

The concept of duality allows us to deduce some properties of information combining bounds when combined with the chain rule shown above. 
\begin{lemma}[Symmetry]\label{Lem:Sym}
Given channels $W_1$ and $W_2$, there exist channels $\bar W_1$ and $\bar W_2$ with $H(\bar W_i) = \log2 - H(W_i)$ such that 
\begin{align}
&\widetilde H^\downarrow_\alpha(W_1\boxast W_2) - \frac12\left( \widetilde H^\downarrow_\alpha(W_1) + \widetilde H^\downarrow_\alpha(W_2)\right)  \nonumber\\
&= \widetilde H^\downarrow_\alpha(\bar W_1\boxast \bar W_2) - \frac12\left( \widetilde H^\downarrow_\alpha(\bar W_1) + \widetilde H^\downarrow_\alpha(\bar W_2)\right) .
\end{align}
\end{lemma}
It follows that the optimal lower and upper bounds have to necessarily be symmetric. This property was known for $\alpha=1$ and the new chain rule provides the missing link in extending this result to \Renyi entropies. The interplay between symmetry, duality and chain rule is visualized in Figure~\ref{Fig:relations}.

\subsection{Special case: $\alpha=2$}

Based on the classical result and the symmetry property of the quantum problem it appears that the solution on the case of $\alpha=2$ should be particularly simple. We will see now that this is indeed the case. 

It is well known that 
\begin{align}
\widetilde H_2^\downarrow(X|B) = -\log \Ppg(X|B), 
\end{align}
where $\Ppg(X|B)$ is the guessing probability using the so-called pretty good measurement, 
\begin{align}
\Ppg(X|B) = \sum_x p(x) \tr \rho_x M_x, 
\end{align}
with
\begin{align}
M_x = \rho_B^{-\frac12} p(x) \rho_x \rho_B^{-\frac12}. 
\end{align}
We can now prove the following. 
\begin{proposition}\label{Prop:H2down}
We have,
\begin{align}
&\widetilde H_2^\downarrow(X_1 + X_2|B_1B_2)_\tau \nonumber\\
&= - \log \left[ 2 \left(\frac12 - e^{-\widetilde H_2^\downarrow(X_1|B_1)_{\rho_1}}\right)\left(\frac12 - e^{-\widetilde H_2^\downarrow(X_2|B_2)_{\rho_2}}\right) + \frac12\right]. 
\end{align}
\end{proposition}
%\begin{proof}
%Follows by direct calculation. For additional details, see Appendix~\ref{Appendix:Proofs}.  
%\end{proof}
This result resembles nicely the result in the classical case, compare~\cite[Theorem IV.9]{hirche2020renyi}, and just as in the classical case equality holds~\cite[Lemma IV.11]{hirche2020renyi}. 

By the chain rule, this also implies the following. 
\begin{proposition}\label{Prop:H2up}
    We have, 
    \begin{align}
        &\widebar H_\frac12^\uparrow(X_2|X_1+X_2,B_1B_2)_\tau \nonumber\\
        &= \log\left( 1+ \left(e^{\widebar H_\frac12^\uparrow(X_1|B_1)_{\rho_1}}-1\right)\left(e^{\widebar H_\frac12^\uparrow(X_2|B_2)_{\rho_2}}-1\right)\right). 
    \end{align}
\end{proposition}
We remark that an equality expressions such as the ones above should also hold in the case of $\alpha=3$, however calculating it explicitly seems more difficult.

\subsection{Conjecture for general $\alpha$}\label{Sec:Conjectures}

The following conjectures generalize those made in~\cite[Conjecture VII.1]{hirche2018bounds} and~\cite[Conjecture VII.2]{hirche2018bounds} to \Renyi entropies. 

 \begin{conjecture}[$\widetilde H_\alpha^\downarrow$ BSC-PSC bound]
Let $H_1=\widetilde H_\alpha^\downarrow(X_1|B_1)_{\rho_1}$ and $H_2=\widetilde H_\alpha^\downarrow(X_2|B_2)_{\rho_2}$. For $\alpha\in(0,2]\cup[3,\infty)$, 
 \begin{align}
 &\widetilde H_\alpha^\downarrow(X_1+X_2|B_1B_2)_\tau \nonumber\\
&\geq 
 \begin{cases} h_\alpha(h_\alpha^{-1}( H_1)\ast h_\alpha^{-1}( H_2)) &H_1 + H_2 \leq \log 2 \\
 H_1 + H_2 - \log2 
 + h_\alpha(h_\alpha^{-1}(\log2-H_1)\ast h_\alpha^{-1}( \log2 -H_2)) 
 &H_1 + H_2 \geq \log 2 \end{cases}. \label{BSCdownH}
 \end{align}
For $\alpha\in[2,3]$ the same holds with $\geq$ exchanged by $\leq$ and for $\alpha \in \{2,3\}$ the above holds with equality. 
\end{conjecture}
If true, these bounds are tight in the sense that equality is achieved by binary symmetric and pure state channels respectively. 
\begin{conjecture}[$\widetilde H_\alpha^\downarrow$ BEC-bound]\label{thm:q-BEC-H}
For $\alpha\in(0,2]\cup[3,\infty)$,  
\begin{align}
 &\widetilde H_\alpha^\downarrow(X_1+X_2|B_1B_2)_\tau \nonumber\\
 &\leq \frac{1}{1-\alpha}\log\frac{(\delta_\alpha^H - \widetilde K_\alpha^\downarrow(X_1|B_1))(\delta_\alpha^H - \widetilde K_\alpha^\downarrow(X_2|B_2))}{1-\delta_\alpha^H} + \delta_\alpha^H  .
 \end{align}
For $\alpha\in[2,3]$ the same holds with $\geq$ exchanged by $\leq$ and for $\alpha \in \{2,3\}$ the above holds with equality. 
\end{conjecture}
Again, if true, these bounds are tight in the sense that equality is achieved by binary erasure channels. 
The special case of $\alpha=2$ was discussed in the previous section. The conjectured bounds are symmetrized versions of the classical bounds. From the previous discussion it follows that the optimal bounds must have this symmetry. Finally, the conjecture has withstood numerical testing, see Appendix~\ref{App:Numerics}. 

Note that the conjectured bounds also imply conjectures on bounds for 
\begin{align}
    \widebar H_\alpha^\uparrow(X_2|X_1+X_2,B_1B_2)_\tau
\end{align}
via the previously discussed information combining chain rule.

\section{Conclusions}
In this work we revisited information combining bounds in the classical and quantum setting. The main contributions are new chain rules for \Renyi entropies that connect different definitions of such entropies. In the classical case, these allow us to complete the picture on \Renyi information combining for the Arimoto conditional entropy. Besides a theoretical interest these results might also have practical applications, e.g. towards a better understanding of the channel polarization phenomenon when  considering \Renyi entropies. In the quantum setting, the chain rules allow for an extension of properties such as symmetry to \Renyi entropy and a new conjecture for optimal quantum bounds that extends previous $\alpha=1$ conjectures. The results also provide strong evidence that $\widetilde H^\downarrow_\alpha$ is the \Renyi entropy of choice when generalizing information combining bounds. 

\section*{Acknowledgments} 
This project has received funding from the European Union's Horizon 2020 research and innovation programme under the Marie Sklodowska-Curie Grant Agreement No. H2020-MSCA-IF-2020-101025848. This research is supported by the National Research Foundation, Singapore and A*STAR under its CQT Bridging Grant.

\bibliographystyle{abbrv}
\bibliography{library}

\appendices

\section{Numerical Results}\label{App:Numerics}
In this section we will discuss some numerical results concerning the setting with quantum side information. In all cases we fix $W_1=W_2$ and plot the difference 
\begin{align}
H_\alpha^\star(X_1+X_2|B_1B_2) - H_\alpha^\star(X_1|B_2)
\end{align}
against the initial entropy $H_\alpha^\star(X_1|B_2)$. The lines give the bounds evaluated for the binary erasure channel (BEC, green), binary symmetric channel (BSC, blue) and the pure state channel (PSC, red). Additionally, we plotted 1000 points based on randomly chosen channels with the $B_i$ being 2-dimensional. 
\begin{figure*}
    \centering
    \includegraphics[width=0.95\textwidth]{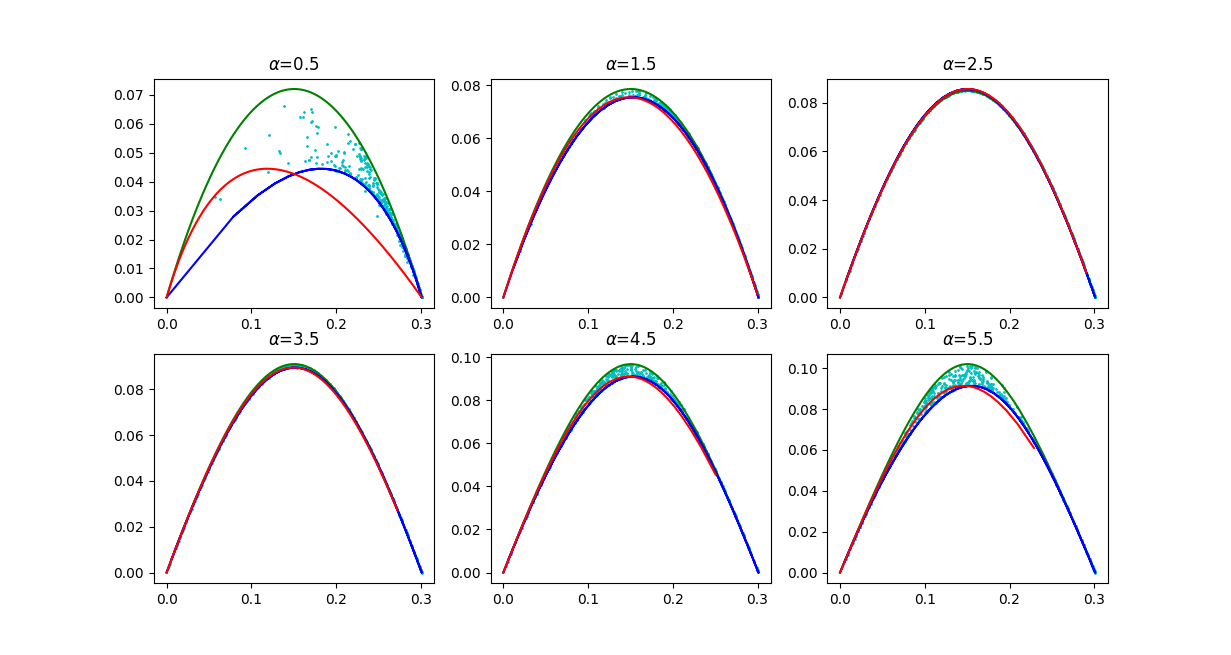}
    \caption{Numerical check of bounds for $\widetilde H_\alpha^\downarrow$ for different values of $\alpha$ with randomly drawn states. The conjectures stated in Section~\ref{Sec:Conjectures} seem to hold.}
    \label{fig:Test1}
\end{figure*}
First, we consider the case of $\widetilde H_\alpha^\downarrow$ in Figure~\ref{fig:Test1} which is the focus of the main text. One immediately verifies the symmetry proven earlier. Additionally, the randomly chosen channels appear to support the conjectures made in Section~\ref{Sec:Conjectures}. 

\begin{figure*}
    \centering
    \includegraphics[width=0.95\textwidth]{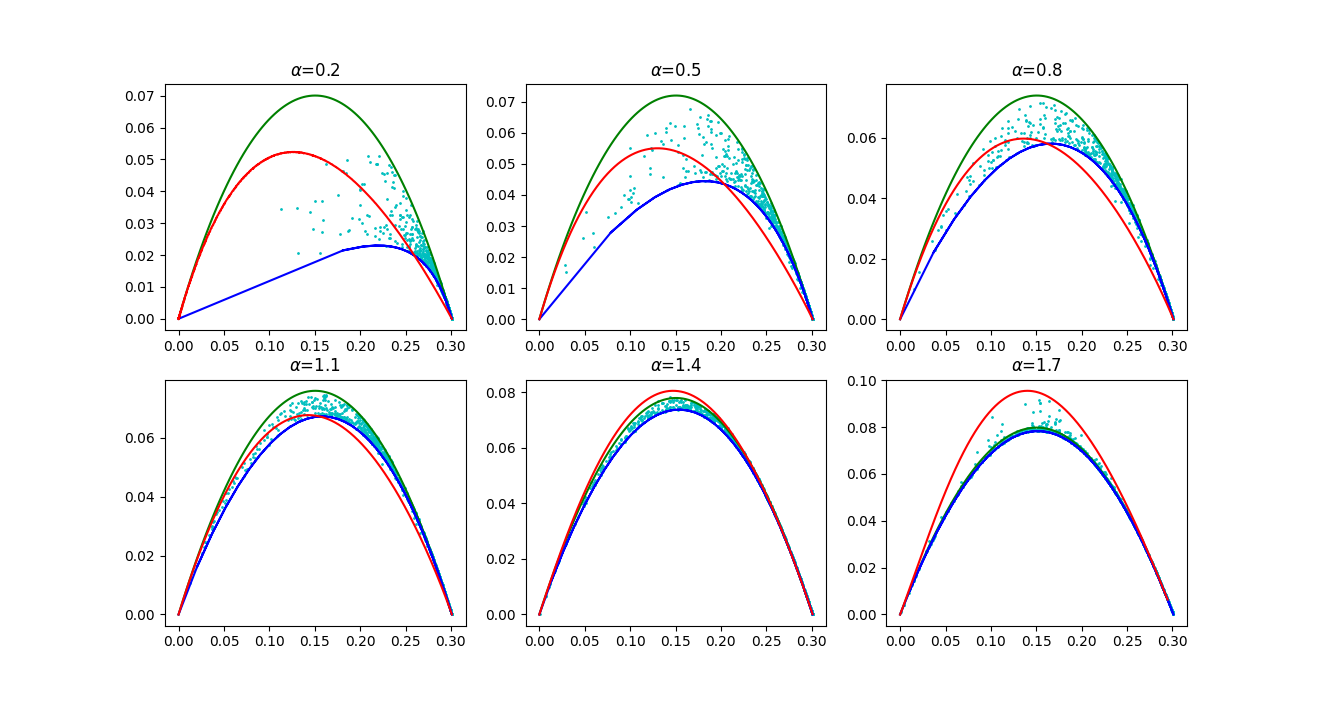}
    \caption{Numerical check for $\widebar H^\downarrow_\alpha$. We clearly do not observe the same symmetry present for $\alpha=1$.  Nevertheless, outer bounds based on the plotted channels seem to hold.}
    \label{fig:Test2}
\end{figure*}
Next we consider the case of $\widebar H^\downarrow_\alpha$ in Figure~\ref{fig:Test2}. We first notice that in this case the analog of the previously observed symmetry property does not seem to hold. Additionally, we notice that while before the BSC and PSC seemed to jointly give a lower or upper bound, in this case the order is less clear. In particular at $\alpha=1.4$ it can be observed that the PSC appears to give an upper bound and the BSC a lower bound, with the BEC sandwiched in between. In summary, while BSC, BEC and PSC still seem to give the outer bounds, many of the properties of the $\alpha=1$ case seem to no longer hold. 

\begin{figure*}
    \centering
    \includegraphics[width=0.95\textwidth]{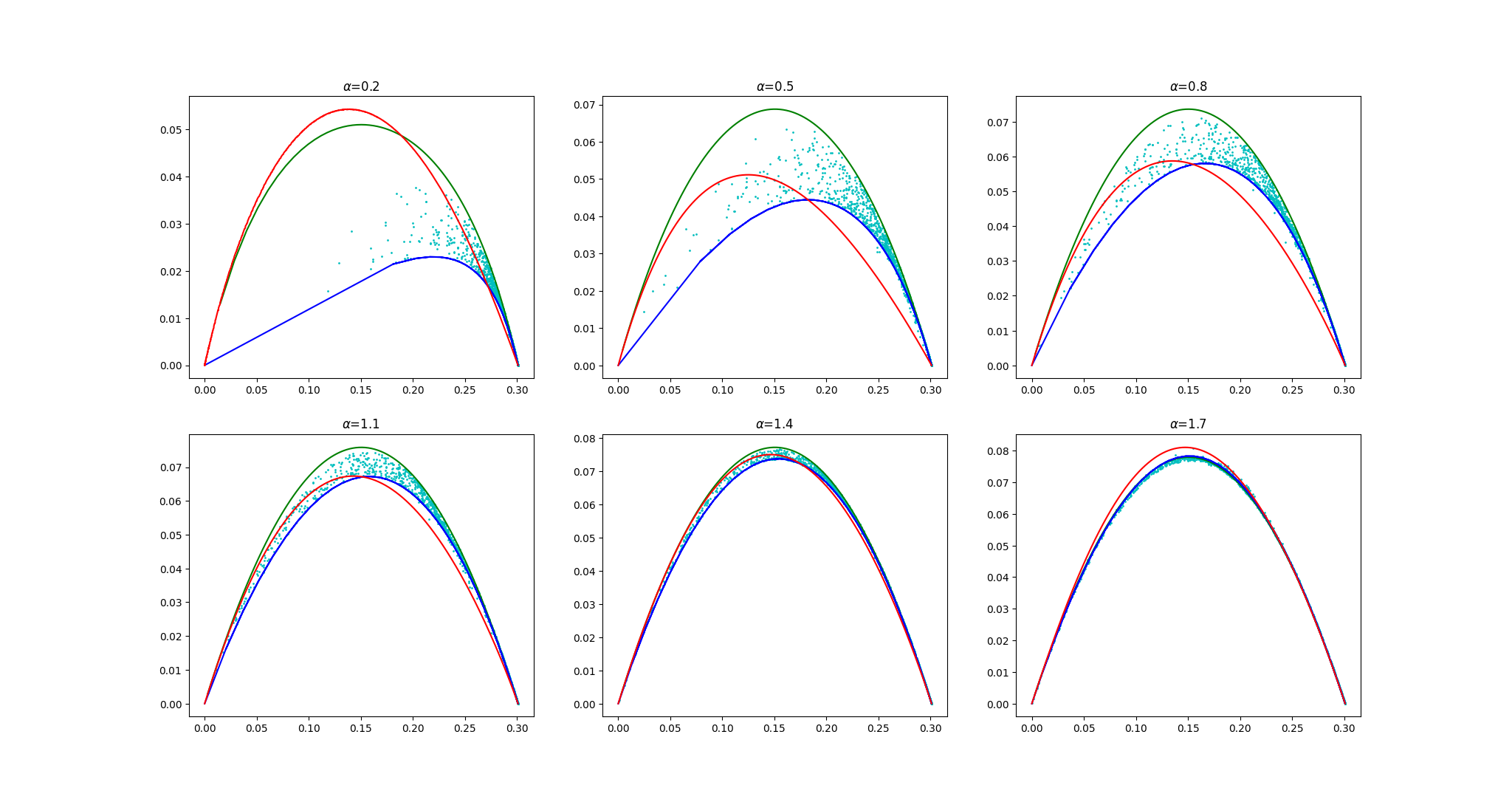}
    \caption{Numerical check for $\widebar H^\uparrow_\alpha$. Again, we clearly do not observe the same symmetry present for $\alpha=1$. Outer bounds are violated for certain $\alpha$ but that's expected, see main text.}
    \label{fig:Test3}
\end{figure*}
Finally, we consider $\widebar H^\uparrow_\alpha$ in Figure~\ref{fig:Test3}. Again we see that the desired symmetry property does not seem to hold. Additionally, we observe that the considered channels do not seem to give outer bounds any longer, see $\alpha=1.7$. Note however that even in the classical case the usual information combining bounds do not hold for this $\alpha$, see~\cite{hirche2020renyi} for details. Therefore the examples observed here are somewhat to be expected. 

In summary, we see that while many of the properties observed for $\alpha=1$ still hold when considering $\widetilde H_\alpha^\downarrow$ this is not true for other choices of \Renyi entropies. This seems to imply that $\widetilde H_\alpha^\downarrow$ should be the \Renyi entropy of choice when generalizing information combining bounds. 

\section{Additional proofs} \label{Appendix:Proofs}
In this appendix we provide the proofs omitted in the main text. For the convenience of the reader we repeat the statements of the individual results before proving them. 
\newtheorem*{lem:cl-CR}{Lemma~\ref{lem:cl-CR}}
\begin{lem:cl-CR}[Classical \Renyi chain rule]
Let $p(x,y)$ and $\bar p(x,y)$ be joint probability distributions related by, 
\begin{align}
\bar p(x,y) = \frac{p(x|y)^\alpha p(y) }{\sum_{x,y} p(x|y)^\alpha p(y)}, \label{Eq:Pxy-relation}
\end{align}
then 
\begin{align}
H^H_\alpha(X|Y)_p = H^A_{\frac{1}{\alpha}}(X|Y)_{\bar p}. 
\end{align}
Similarly, let 
\begin{align}
\hat p(x,y,z) = \frac{p(z) p(x,y|z)^\alpha}{\sum_{x,y,z}p(z) p(x,y|z)^\alpha }, 
\end{align}
then
\begin{align}
H^H_\alpha(XY|Z)_p &= H^A_{\frac{1}{\alpha}}(X|YZ)_{\hat p} + H^H_\alpha(Y|Z)_p \\
&= H^A_{\frac{1}{\alpha}}(XY|Z)_{\hat p}.
\end{align}
\end{lem:cl-CR}
\begin{proof}
We show the first equality in the last statement. All the others can similarly be checked by substitution.
\begin{align}
&H^A_{\frac{1}{\alpha}}(X|YZ)_{\hat p} \\
&= \frac{1}{\alpha-1}\log\sum_y \hat p(y) \left(\sum_x \hat p(x|y)^\frac{1}{\alpha}\right)^{\alpha} \\
&= \frac{1}{\alpha-1}\log\sum_y  \left(\sum_x \hat p(x,y)^\frac{1}{\alpha}\right)^{\alpha} \\
&= \frac{1}{\alpha-1}\log\sum_y  \left(\sum_x \left(\frac{p(z) p(x,y|z)^\alpha}{\sum_{x,y,z}p(z) p(x,y|z)^\alpha }\right)^\frac{1}{\alpha}\right)^{\alpha} \\
&= \frac{1}{\alpha-1}\log\sum_y  \left(\sum_x \left(p(z) p(x,y|z)^\alpha\right)^\frac{1}{\alpha}\right)^{\alpha} \nonumber\\ &\quad- \frac{1}{\alpha-1}\log\sum_{x,y,z}p(z) p(x,y|z)^\alpha  \\
&= \frac{1}{\alpha-1}\log\sum_y  \left( p(z)^\frac{1}{\alpha} p(y|z)\right)^{\alpha} + H^H_\alpha(X, Y|Z)_p  \\
&= - H^H_\alpha(Y|Z)_p + H^H_\alpha(X, Y|Z)_p \, .
\end{align}
\end{proof}
\newtheorem*{Lem:CICCR}{Lemma~\ref{Lem:CICCR}}
\begin{Lem:CICCR}[Information combining chain rule]
Given channels $W_1$ and $W_2$, there exist channels $W'_1$ and $W'_2$ with $H^H_\alpha(W_i) =  H^A_{\frac{1}{\alpha}}(W'_i)$ such that 
\begin{align}
H^H_\alpha(W_1) + H^H_\alpha(W_2) = H^A_{\frac{1}{\alpha}}(W'_1\varoast W'_2) + H^H_\alpha(W_1\boxast W_2).
\end{align}
More precisely, these channels are given by
\begin{align}
W'_i\sim \bar p_i(x_i,y_i) = \frac{p_i(x_i|y_i)^\alpha p_i(y_i) }{\sum_{x,y} p_i(x_i|y_i)^\alpha p_i(y_i)}.
\end{align}
\end{Lem:CICCR}
\begin{proof}
    This follows essentially from the chain rule in Lemma~\ref{lem:cl-CR} in the previous section. What remains to be shown is that 
    \begin{align}
H^A_{\frac{1}{\alpha}}(A|BC)_{\hat p} = H^A_{\frac{1}{\alpha}}(W'_1\varoast W'_2),
    \end{align}
i.e. that $\hat p$ is a valid combined channel. 
    This can be seen by direct calculation, compare also the quantum case later on. 
\end{proof}
\newtheorem*{Thm:AclUpper}{Theorem~\ref{Thm:AclUpper}}
\begin{Thm:AclUpper}
    For $\alpha\in(0,\frac13]\cup[\frac12,\infty)$, it holds that
    \begin{align}
    &H^A_{\alpha}(X_2|X_1+X_2,Y_1Y_2) \nonumber\\
    &\leq H^A_{\alpha}(X_1|Y_1) + H^A_{\alpha}(X_2|Y_2) - h_{\frac1{\alpha}}(h_{\frac1{\alpha}}^{-1}( H_{\alpha}^A(X_1|Y_1))\ast h_{\frac1{\alpha}}^{-1}( H_{\alpha}^A(X_2|Y_2))).
    \end{align}
    For $\alpha\in[\frac13,\frac12]$ the same holds with $\leq$ exchanged by $\geq$. Note that for $\alpha \in \{\frac13,\frac12 \}$ the above holds with equality. These bounds are tight in the sense that equality is achieved by binary symmetric channels. 
\end{Thm:AclUpper}
\begin{proof}
Using the chain rule, we find
    \begin{align}
    &H^A_{\alpha}(X_2|X_1+X_2,Y_1Y_2) \nonumber\\
    &= H^A_{\alpha}(W_1\varoast W_2)  \\ 
    &= H^A_{\alpha}(W_1) + H^A_{\alpha}(W_2) - H^H_{\frac1{\alpha}}(W'_1\boxast W'_2).  
    \end{align}
    From here, if 
 $\alpha\in(0,\frac13]\cup[\frac12,\infty)$ then by using Theorem~\ref{thm:old-H-BSC}, 
    \begin{align}
     &H^A_{\alpha}(X_2|X_1+X_2,Y_1Y_2) - H^A_{\alpha}(W_1) - H^A_{\alpha}(W_2)  \\
    &\quad \leq - h_{\frac1{\alpha}}(h_{\frac1{\alpha}}^{-1}( H_{\frac1{\alpha}}^H(W'_1))\ast h_{\frac1{\alpha}}^{-1}( H_{\frac1{\alpha}}^H(W'_2))) \\
    &\quad = - h_{\frac1{\alpha}}(h_{\frac1{\alpha}}^{-1}( H_{\alpha}^A(W_1))\ast h_{\frac1{\alpha}}^{-1}( H_{\alpha}^A(W_2)))
    \end{align}
    and otherwise with the initial inequality turned around. 
\end{proof}

\newtheorem*{Thm:AclLower}{Theorem~\ref{Thm:AclLower}}
\begin{Thm:AclLower}
   For $\alpha\in(0,\frac13]\cup[\frac12,\infty)$, it holds that
    \begin{align}
        &H^A_{\alpha}(X_2|X_1+X_2,Y_1Y_2) \\
        &\geq H^A_{\alpha}(X_1|Y_1) + H^A_{\alpha}(X_2|Y_2) - \frac{\alpha}{\alpha-1}\log\frac{(\delta_{\frac1{\alpha}} - K_{\frac1{\alpha}}^A(X_1|Y_1))(\delta_{\frac1{\alpha}} - K_{\frac1{\alpha}}^A(X_2|Y_2))}{1-\delta_{\frac1{\alpha}}} + \delta_{\frac1{\alpha}}.
    \end{align}
    For $\alpha\in[\frac13,\frac12]$ the same holds with $\geq$ exchanged by $\leq$. Note that for $\alpha \in \{\frac13,\frac12 \}$ the above holds with equality. These bounds are tight in the sense that equality is achieved by binary erasure channels. 
\end{Thm:AclLower}
\begin{proof}
Starting similar as before, we get that, if $\alpha\in(0,\frac13]\cup[\frac12,\infty)$ then by Theorem~\ref{thm:old-H-BEC}, 
    \begin{align}
    &H^A_{\alpha}(X_2|X_1+X_2,Y_1Y_2) - H^A_{\alpha}(W_1) - H^A_{\alpha}(W_2)  \\    
    &\geq - \frac{\alpha}{\alpha-1}\log\frac{(2^{\frac{\alpha-1}{\alpha}} - e^{\frac{\alpha-1}{\alpha}H^H_{\frac{1}{\alpha}}(W'_1)})(2^{\frac{\alpha-1}{\alpha}} - e^{\frac{\alpha-1}{\alpha}H^H_{\frac{1}{\alpha}}(W'_2)})}{1-2^{\frac{\alpha-1}{\alpha}}} + 2^{\frac{\alpha-1}{\alpha}} \\
    &= - \frac{\alpha}{\alpha-1}\log\frac{(2^{\frac{\alpha-1}{\alpha}} - e^{\frac{\alpha-1}{\alpha}H^A_{\alpha}(W_1)})(2^{\frac{\alpha-1}{\alpha}} - e^{\frac{\alpha-1}{\alpha}H^A_{\alpha}(W_2)})}{1-2^{\frac{\alpha-1}{\alpha}}} + 2^{\frac{\alpha-1}{\alpha}}
    \end{align}
    and otherwise with the inequalities turned around. 
\end{proof}

\newtheorem*{Lem:QCR}{Lemma~\ref{Lem:QCR}}
\begin{Lem:QCR}[Quantum \Renyi chain rule]
Let $\rho_{AB}$ and $\bar\rho_{AB}$ be quantum states related by
\begin{align}
\bar\rho_{AB} = \frac{\left(\rho_B^{\frac{1-\alpha}{2\alpha}}\rho_{AB}\rho_B^{\frac{1-\alpha}{2\alpha}}\right)^\alpha}{\tr\left(\rho_B^{\frac{1-\alpha}{2\alpha}}\rho_{AB}\rho_B^{\frac{1-\alpha}{2\alpha}}\right)^\alpha}, 
\end{align}
then
\begin{align}
\widetilde H^\downarrow_\alpha(A|B)_\rho = \widebar H^\uparrow_{\frac{1}{\alpha}}(A|B)_{\bar\rho}.
\end{align}
Furthermore, if 
\begin{align}
\widetilde\rho_{ABC} = \frac{\left(\rho_C^{\frac{1-\alpha}{2\alpha}}\rho_{ABC}\rho_C^{\frac{1-\alpha}{2\alpha}}\right)^\alpha}{\tr\left(\rho_C^{\frac{1-\alpha}{2\alpha}}\rho_{ABC}\rho_C^{\frac{1-\alpha}{2\alpha}}\right)^\alpha}, 
\end{align}
then
\begin{align}
\widetilde H^\downarrow_\alpha(AB|C)_\rho = \widebar H^\uparrow_{\frac{1}{\alpha}}(A|BC)_{\widetilde\rho} + \widetilde H^\downarrow_\alpha(B|C)_\rho = \widebar H^\uparrow_{\frac{1}{\alpha}}(AB|C)_{\widetilde\rho}.
\end{align}
\end{Lem:QCR}
\begin{proof}
Again, we show one and the others are similar. 
\begin{align}
&\widebar H^\uparrow_{\frac{1}{\alpha}}(A|BC)_{\widetilde\rho} \nonumber\\
&= \frac{1}{\alpha-1}\log\tr\left( \left(\tr_A\widetilde\rho_{ABC}^{\frac{1}{\alpha}}\right)^\alpha\right) \\
&= \widetilde H^\downarrow_\alpha(AB|C)_\rho + \frac{1}{\alpha-1}\log\tr\left( \tr_A\rho_C^{\frac{1-\alpha}{2\alpha}}\rho_{ABC}\rho_C^{\frac{1-\alpha}{2\alpha}}\right)^\alpha\\
&= \widetilde H^\downarrow_\alpha(AB|C)_\rho + \frac{1}{\alpha-1}\log\tr\left( \rho_C^{\frac{1-\alpha}{2\alpha}}\rho_{BC}\rho_C^{\frac{1-\alpha}{2\alpha}}\right)^\alpha\\
&= \widetilde H^\downarrow_\alpha(AB|C)_\rho - \widetilde H^\downarrow_\alpha(B|C)_\rho.
\end{align}
\end{proof}

\newtheorem*{Lem:QICCR}{Lemma~\ref{Lem:QICCR}}
\begin{Lem:QICCR}[Information combining chain rule]
Given channels $W_1$ and $W_2$, there exist channels $W'_1$ and $W'_2$ with $\widetilde H^\downarrow_\alpha(W_i) = \widebar H^\uparrow_{\frac{1}{\alpha}}(W'_i)$ such that 
\begin{align}
\widetilde H^\downarrow_\alpha(W_1) + \widetilde H^\downarrow_\alpha(W_2) = \widebar H^\uparrow_{\frac{1}{\alpha}}(W'_1\varoast W'_2) + \widetilde H^\downarrow_\alpha(W_1\boxast W_2).
\end{align}
More precisely, these channels are given by
\begin{align}
W'_i\sim \bar\rho_{X_iB_i} = \frac{\left(\rho_{B_i}^{\frac{1-\alpha}{2\alpha}}\rho_{X_iB_i}\rho_{B_i}^{\frac{1-\alpha}{2\alpha}}\right)^\alpha}{\tr\left(\rho_{B_i}^{\frac{1-\alpha}{2\alpha}}\rho_{X_iB_i}\rho_{B_i}^{\frac{1-\alpha}{2\alpha}}\right)^\alpha}.
\end{align}
\end{Lem:QICCR}
\begin{proof}
From the previous section we know that the given $W'_i$ are such that $\widetilde H^\downarrow_\alpha(W_i) = \widebar H^\uparrow_{\frac{1}{\alpha}}(W'_i)$ and furthermore we get from the chain rule in Lemma~\ref{Lem:QCR} that 
\begin{align}
\widetilde H^\downarrow_\alpha(W_1) + \widetilde H^\downarrow_\alpha(W_2) = \widebar H^\uparrow_{\frac{1}{\alpha}}(A|BC)_{\widetilde\tau} + \widetilde H^\downarrow_\alpha(W_1\boxast W_2).
\end{align}
It remains to show that $\widetilde\tau$ corresponds to a CNOT of $W'_1$ and $W'_2$. This can be shown by direct calculation using the block structure of classical-quantum states. In short one can check that 
\begin{align}
\bar\rho_{X_iB_i} = \frac12 |0\rangle\langle 0| \otimes \bar\sigma_0^{B_i} + \frac12 |1\rangle\langle 1| \otimes \bar\sigma_1^{B_i},
\end{align}
where $\bar\sigma_i^{B_i} = \frac{\left((\sigma_0^{B_i}+ \sigma_1^{B_i})^{\frac{1-\alpha}{2\alpha}}\sigma_i^{B_i}(\sigma_0^{B_i}+ \sigma_1^{B_i})^{\frac{1-\alpha}{2\alpha}}\right)^\alpha}{\tr\left((\sigma_0^{B_i}+ \sigma_1^{B_i})^{\frac{1-\alpha}{2\alpha}}\sigma_i^{B_i}(\sigma_0^{B_i}+ \sigma_1^{B_i})^{\frac{1-\alpha}{2\alpha}}\right)^\alpha}$ and 
\begin{align}
   \widetilde\tau  = \sum_{z,x_2} \frac14 |z+x_2\rangle\langle z+x_2| \otimes |x_2\rangle\langle x_2| \otimes \bar\sigma^{B_1}_z \otimes \bar\sigma^{B_2}_{x_2}. 
\end{align}
From there it immediately follows that
\begin{align}
\widebar H^\uparrow_{\frac{1}{\alpha}}(A|BC)_{\widetilde\tau} = \widebar H^\uparrow_{\frac{1}{\alpha}}(W'_1\varoast W'_2),
\end{align}
which concludes the proof. 
\end{proof}

\newtheorem*{Lem:Sym}{Lemma~\ref{Lem:Sym}}
\begin{Lem:Sym}[Symmetry]
Given channels $W_1$ and $W_2$, there exist channels $\bar W_1$ and $\bar W_2$ with $H(\bar W_i) = \log2 - H(W_i)$ such that 
\begin{align}
&\widetilde H^\downarrow_\alpha(W_1\boxast W_2) - \frac12\left( \widetilde H^\downarrow_\alpha(W_1) + \widetilde H^\downarrow_\alpha(W_2)\right)  \nonumber\\
&= \widetilde H^\downarrow_\alpha(\bar W_1\boxast \bar W_2) - \frac12\left( \widetilde H^\downarrow_\alpha(\bar W_1) + \widetilde H^\downarrow_\alpha(\bar W_2)\right) .
\end{align}
\end{Lem:Sym}
\begin{proof}
Starting from the chain rule we get
\begin{align}
&\widetilde H^\downarrow_\alpha(W_1\boxast W_2) - \frac12\left(\widetilde H^\downarrow_\alpha(W_1) + \widetilde H^\downarrow_\alpha(W_2)\right) \nonumber\\
&= \frac12\left(\widetilde H^\downarrow_\alpha(W_1) + \widetilde H^\downarrow_\alpha(W_2)\right) - \widebar H^\uparrow_{\frac{1}{\alpha}}(W'_1\varoast W'_2)  \\
&= \frac12\left(\widebar H^\uparrow_{\frac{1}{\alpha}}(W'_1) + \widebar H^\uparrow_{\frac{1}{\alpha}}(W'_2)\right) - \widebar H^\uparrow_{\frac{1}{\alpha}}(W'_1\varoast W'_2)  \\
&= \widetilde H^\downarrow_\alpha( W'^\bot_1\boxast \bar W'^\bot_2) - \frac12\left( \widetilde H^\downarrow_\alpha(\bar W'^\bot_1) + \widetilde H^\downarrow_\alpha(\bar W'^\bot_2)\right),
\end{align}
from where the claim follows with $\bar W_1=W'^\bot_1$ and $\bar W_2=W'^\bot_2$. 
\end{proof}

\newtheorem*{Prop:H2down}{Proposition~\ref{Prop:H2down}}
\begin{Prop:H2down}
We have,
\begin{align}
&\widetilde H_2^\downarrow(X_1 + X_2|B_1B_2)_\tau \nonumber\\
&= - \log \left[ 2 \left(\frac12 - e^{-\widetilde H_2^\downarrow(X_1|B_1)_{\rho_1}}\right)\left(\frac12 - e^{-\widetilde H_2^\downarrow(X_2|B_2)_{\rho_2}}\right) + \frac12\right]. 
\end{align}
\end{Prop:H2down}
\begin{proof}
By rather direct calculation, it can be checked that, 
\begin{align}
&\Ppg(X_1 + X_2|B_1B_2) \nonumber\\
&= \sum_{i,j}\tr(\sigma^{B_1}_i \otimes \sigma^{B_2}_{i})(M_j\otimes M_j) + \sum_{i,j}\tr(\sigma^{B_1}_i \otimes \sigma^{B_2}_{i\oplus 1})(M_j\otimes M_{j\oplus 1}) \\
&= \Ppg(X_1|B_1)\Ppg(X_2|B_2) + (1-\Ppg(X_1|B_1))(1-\Ppg(X_2|B_2)) \\
&= 2 (\frac12 - \Ppg(X_1|B_1))(\frac12 - \Ppg(X_2|B_2)) + \frac12,
\end{align}
where $M_i=\frac12 \bar\sigma^{-\frac12} \sigma_i \bar\sigma^{-\frac12}$ with $\bar\sigma = \frac12(\sigma_0+\sigma_1)$ are the pretty good measurements with respect to the individual states. The claim follows then by taking $-\log$ of all probabilities. 
\end{proof}

\newtheorem*{Prop:H2up}{Proposition~\ref{Prop:H2up}}
\begin{Prop:H2up}
    We have, 
    \begin{align}
        &\widebar H_\frac12^\uparrow(X_2|X_1+X_2,B_1B_2)_\tau \nonumber\\
        &= \log\left( 1+ \left(e^{\widebar H_\frac12^\uparrow(X_1|B_1)_{\rho_1}}-1\right)\left(e^{\widebar H_\frac12^\uparrow(X_2|B_2)_{\rho_2}}-1\right)\right). 
    \end{align}
\end{Prop:H2up}
\begin{proof}
The above can be checked by direct calculation, noting that 
\begin{align}
    \widebar H_\frac12^\uparrow(X_1|B_1) &= \log(1+ \tr\sqrt{\sigma_0\sigma_1}) \\ 
    \widebar H_\frac12^\uparrow(X_2|X_1+X_2,B_1B_2) &= \log(1+ \tr\sqrt{\sigma_0\sigma_1}\tr\sqrt{\bar\sigma_0\bar\sigma_1}). 
\end{align}
However, in the spirit of this work, we will give a different proof using the chain rule. Note that
\begin{align}
    &\widebar H_\frac12^\uparrow(X_2|X_1+X_2,B_1B_2) \nonumber\\
    &= \widebar H_\frac12^\uparrow(W_1 \varoast W_2) \\
    &= - \tilde H_2^\downarrow(W'_1 \boxast W'_2) + \widebar H_\frac12^\uparrow(W_1) + \widebar H_\frac12^\uparrow(W_2) \\ 
    &= \log \left[ 2 \left(\frac12 - e^{-\widetilde H_2^\downarrow(W'_1)}\right)\left(\frac12 - e^{-\widetilde H_2^\downarrow(W'_2)}\right) + \frac12\right] + \widebar H_\frac12^\uparrow(W_1) + \widebar H_\frac12^\uparrow(W_2) \\
    &= \log \left[ 2 \left(\frac12 - e^{-\widebar H_\frac12^\uparrow(W_1)}\right)\left(\frac12 - e^{-\widebar H_\frac12^\uparrow(W_2)}\right) + \frac12\right] + \log \left( e^{\widebar H_\frac12^\uparrow(W_1) + \widebar H_\frac12^\uparrow(W_2)}\right)  \\
    &= \log\left( 1+ \left(e^{\widebar H_\frac12^\uparrow(X_1|B_1)}-1\right)\left(e^{\widebar H_\frac12^\uparrow(X_2|B_2)}-1\right)\right), 
\end{align}
where we have used the chain rule and the fact that the resulting $W'_i$ are such that $\widetilde H_2^\downarrow(W'_i) = \widebar H_\frac12^\uparrow(W_i)$.
\end{proof}

\end{document}